# Three-dimensional hydrodynamic Bondi-Hoyle accretion

## V. Specific heat ratio 1.01, nearly isothermal flow

**M. Ruffert**[*]

Max-Planck-Institut für Astrophysik, Postfach 1523, D - 85740 Garching, Germany

October 4, 1995

**Abstract.** We investigate the hydrodynamics of three-dimensional classical Bondi-Hoyle accretion. A totally absorbing sphere of different sizes (1, 0.1 and 0.02 accretion radii) moves at different Mach numbers (0.6, 1.4, 3.0 and 10) relative to a homogeneous and slightly perturbed medium, which is taken to be an ideal, nearly isothermal, gas ($\gamma = 1.01$). We examine the influence of Mach number of the flow and size of the accretor upon the physical behaviour of the flow and the accretion rates. The hydrodynamics is modeled by the "Piecewise Parabolic Method" (PPM). The resolution in the vicinity of the accretor is increased by multiply nesting several $32^3$-zone grids around the sphere, each finer grid being a factor of two smaller in zone dimension than the next coarser grid. This allows us to include a coarse model for the surface of the accretor (vacuum sphere) on the finest grid while at the same time evolving the gas on the coarser grids.

For small Mach numbers (0.6 and 1.4) the flow patterns tend towards a steady state, while in the case of supersonic flow (Mach 3 and 10) and small enough accretors, (radius of 0.1 and 0.02 accretion radii) an unstable Mach cone develops, destroying axisymmetry. The shock cones in the supersonic models never clear the surface of the accretors (they are tail shocks, not bow shocks) and the opening angle is smaller (compared to models with larger $\gamma$) especially for the highly supersonic models. The densities in the shock cone is larger for models with smaller $\gamma$. The fluctuations of the accretion rates and flow structures are weaker than in the corresponding models with larger $\gamma$. The hydrodynamic drag of all models with accretor sizes of 0.1 $R_A$ or smaller acts in an accelerating direction, while the gravitational drag is always decelerating and larger than the hydrodynamic drag (thus the net force is decelerating).

## 1. Introduction

This last instalment extends our investigation (started with Ruffert, 1994a, Ruffert & Arnett, 1994, Ruffert 1994b and 1995; henceforth referred to as Papers I to IV, respectively) of the classic Bondi-Hoyle-Lyttleton accretion model. In this scenario a totally absorbing sphere moves with velocity $v_\infty$ relative to a surrounding homogeneous medium of density $\rho_\infty$ and sound speed $c_\infty$. One would like to find the accretion rates of various quantities (mass, angular momentum, etc) as well as the properties of the flow (e.g. distribution of matter and velocity, stability, etc). In this fifth instalment we investigate the accretion flow and accreted quantities for a nearly isothermal gas, which we modeled by taking the adiabatic index $\gamma$ to be 1.01 and otherwise using the same parameters as had been done in the simulations of previous papers (which had $\gamma = 5/3$ and $\gamma = 4/3$). In an astrophysical context, a gas might be accreted isothermally when the optical depth is small and the surplus energy (higher temperatures are generated due to gravitative compression of the medium) is radiated away. One of the situations where this might be applicable is the accretion flow behind young stars moving relative to the ambient medium (M. Casali, private communication). Of course a simulation including radiative transport would be better, but it is numerically prohibitive.

The Bondi-Hoyle accretion flow for specific heats $\gamma$ close to unity has been modeled previously by other authors, however most of the simulations were two-dimensional. Shima et al. (1985) find that for $\gamma=1.1$ the shock is attached to the accretor, the opening angle is smaller and the maximum densities are higher, than in cases with larger $\gamma$, and an accretion column is formed. All these two-dimensional results have been reproduced and confirmed by our three-dimensional simulations. Petrich et al. (1989) extended the two-dimensional simulations also including a fully relativistic treatment. Within a factor of two, their results for the mass and momentum accretion rates coincide with the values of Shima et al. (1985) and from the few contour plots shown, also the density distribution seems similar. Matsuda et al. (1992) performed several two-dimensional calculations with $\gamma$ close to one and found that at first radial instability oscillations appear. The numerical resolution of the accretor in these simulations was, however, very crude. It was improved only slightly by Ishii et al. (1993), who nonetheless attempted to additionally simulate three-dimensional isothermal flows. The latter authors conclude that also isothermal flows exhibit the flip-flop instability, but that in 3D it is much less violent than in 2D.

[*] e-mail: `mruffert@mpa-garching.mpg.de`

To more clearly delineate these differences, to compare the accretion rates of several quantities when flow parameters are changed and to obtain a more uniform picture of the accretion rates is the aim of this paper. In section 2 we give only a short summary of the numerical procedure used, since it has been described in the previous Papers I, II and III. Section 3 presents the results, which we analyze and interpret in Sect. 4. Section 5 summarizes the implications of this work.

## 2. Numerical Procedure and Initial Conditions

Since the numerical procedures and initial conditions are mostly identical to what has already been described and used in Papers I to IV we will refrain from repeating every detail. Instead we only give a brief summary.

### 2.1. Numerical Procedure

A gravitating, totally absorbing "sphere" moves through an initially slightly perturbed (3%) but otherwise homogeneous medium. The relative velocity $v_\infty$ is varied in different models: we do simulations with Mach numbers $\mathcal{M}_\infty$ of 0.6, 1.4, 3.0 and 10. In the reference frame of the accretor the surrounding matter flows in +x-direction. Our units are (1) the sound speed $c_\infty$ as velocity unit; (2) the accretion radius (Eq. 3) as unit of length, and (3) $\rho_\infty$ as density unit. Thus the unit of time is $R_A/c_\infty$.

The distribution of matter is discretised on multiply nested (e.g. Berger & Colella, 1989) equidistant Cartesian grids with zone size $\delta$ and is evolved using the "Piecewise Parabolic Method" (PPM) of Colella & Woodward (1984). The equation of state is that of a perfect gas with a specific heat ratio of $\gamma = 1.01$. The model of the maximally accreting, vacuum sphere in a softened gravitational potential is summarized in Paper III.

### 2.2. Models

The combination of parameters that we varied, together with some results are summarised in Table 1. The first letter in the model designation indicates the Mach number: E, F, G and H stand for Mach numbers of 0.6, 1.4, 3.0 and 10, respectively. The second letter specifies the size of the accretor: L (large), M (medium) and S (small) stand for accretor radii of 1, 0.1, and 0.02 $R_A$, respectively. It is informative to compare the sizes analytically for spherically symmetric accretion (Bondi, 1952):

$$d_s = \frac{5 - 3\gamma}{4} R_B \quad , \tag{1}$$

with the Bondi radius given by

$$R_B = \frac{GM}{c_\infty^2} \quad . \tag{2}$$

For further reference we define the accretion radius (Hoyle & Lyttleton, 1939, 1940a, 1940b, 1940c; Bondi & Hoyle, 1944) as

$$R_A = \frac{2GM}{v_\infty^2} \quad , \tag{3}$$

in which $M$ is the mass of the accretor, $G$ the gravitational constant, $c_\infty$ the sound speed of the medium at infinity and $v_\infty$ the relative bulk flow of the medium at infinity. For a value of $\gamma$ of 4/3 (as in Paper IV) the sonic distance is $d_s(\gamma = 4/3) = 0.25 R_B$, while for $\gamma = 1.01$ follows $d_s(\gamma = 1.01) = 0.49 R_B$. Neglecting effects due to the relative motion of medium and accretor and assuming a roughly radial flow close to the accretor, one can boldly eliminate $R_B$ in Eq. 1 in favor of $R_A$ from Eq. 3 and one obtains $d_s = R_A \mathcal{M}_\infty^2 (5 - 3\gamma)/8$, where the Mach number is $\mathcal{M}_\infty = v_\infty/c_\infty$. Inserting both values of gamma, and all three Mach values of the supersonic flow (1.6, 3, 10) one finds, that the sonic distance is in all cases larger than the size of the medium ($0.1 R_A$) and small accretor ($0.02 R_A$). Note, that this is only a rough estimate.

The size of the largest grid $L$ is adapted to the size of the accretor, larger accretors needing larger grids because the evolution time scales are longer and thus the gravitative influence reaches further out. The grids are nested to such a depth $g$ that the radius of the accretor $R$ spans several zones on the finest grid and the softening parameter $\epsilon$ is then chosen to be a few zones less than the number of zones that the accretor spans.

As far as computer resources permitted, we aimed at evolving the models for at least as long as it takes the flow to move from the boundary to the position of the accretor which is at the center (crossing time scale). This time is given by $L/2\mathcal{M}_\infty$ and ranges from about 1 to about 100 time units. The actual time $t_f$ that the model is run can be found in Table 1.

The calculations are performed on a Cray-YMP 4/64 and a Cray J90 8/512. They need about 12 MWords of main memory and take approximately 40 CPU-hours per simulated time unit (for the $\delta = 1/64$ models and Mach 10; the $\delta = 1/256$ models take four times as long, etc.; $\delta$ is the size of a zone on the finest grid, see Table 1).

## 3. Description of Results

We will not only describe the results obtained in the new simulations, but also compare these results with those obtained in Papers II, III and IV for models with $\gamma = 5/3$ and $\gamma = 4/3$.

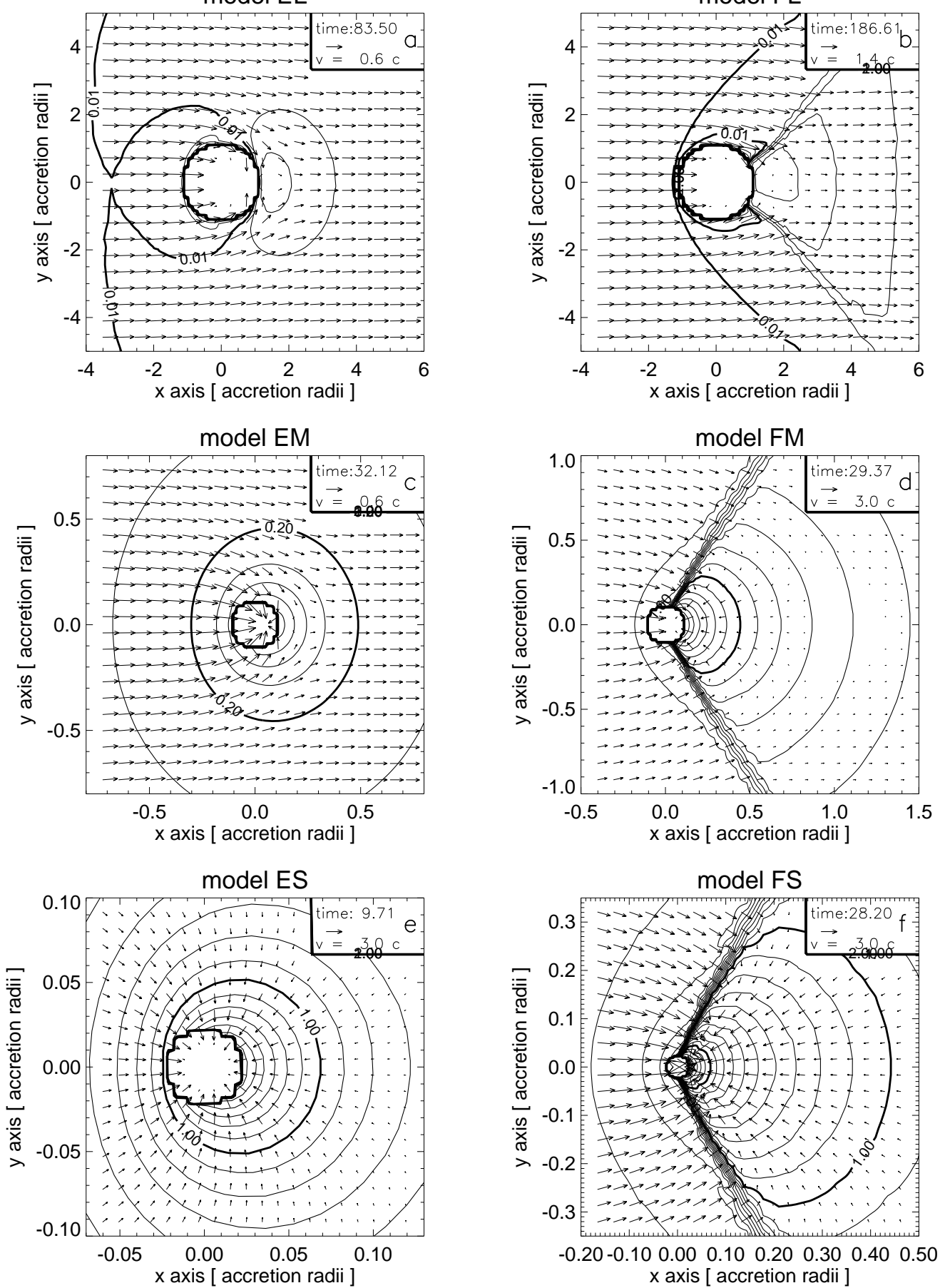

**Fig. 1.** Contour plots showing snapshots of the density together with the flow pattern in a plane containing the center of the accretor for all subsonic models EL (panel a), EM (panel c), ES (panel e) and mildly supersonic models FL (panel b), FM (panel d), FS (panel f). The contour lines are spaced logarithmically in intervals of 0.05 dex for model EL and 0.1 dex for all other models. The bold contour levels correspond to $\log \rho = 0.01$ for model EL and FL, $\log \rho = 0.2$ for model EM, and $\log \rho = 1.0$ for models FM, ES and FS. The time of the snapshot together with the velocity scale is given in the legend in the upper right hand corner of each three panels.

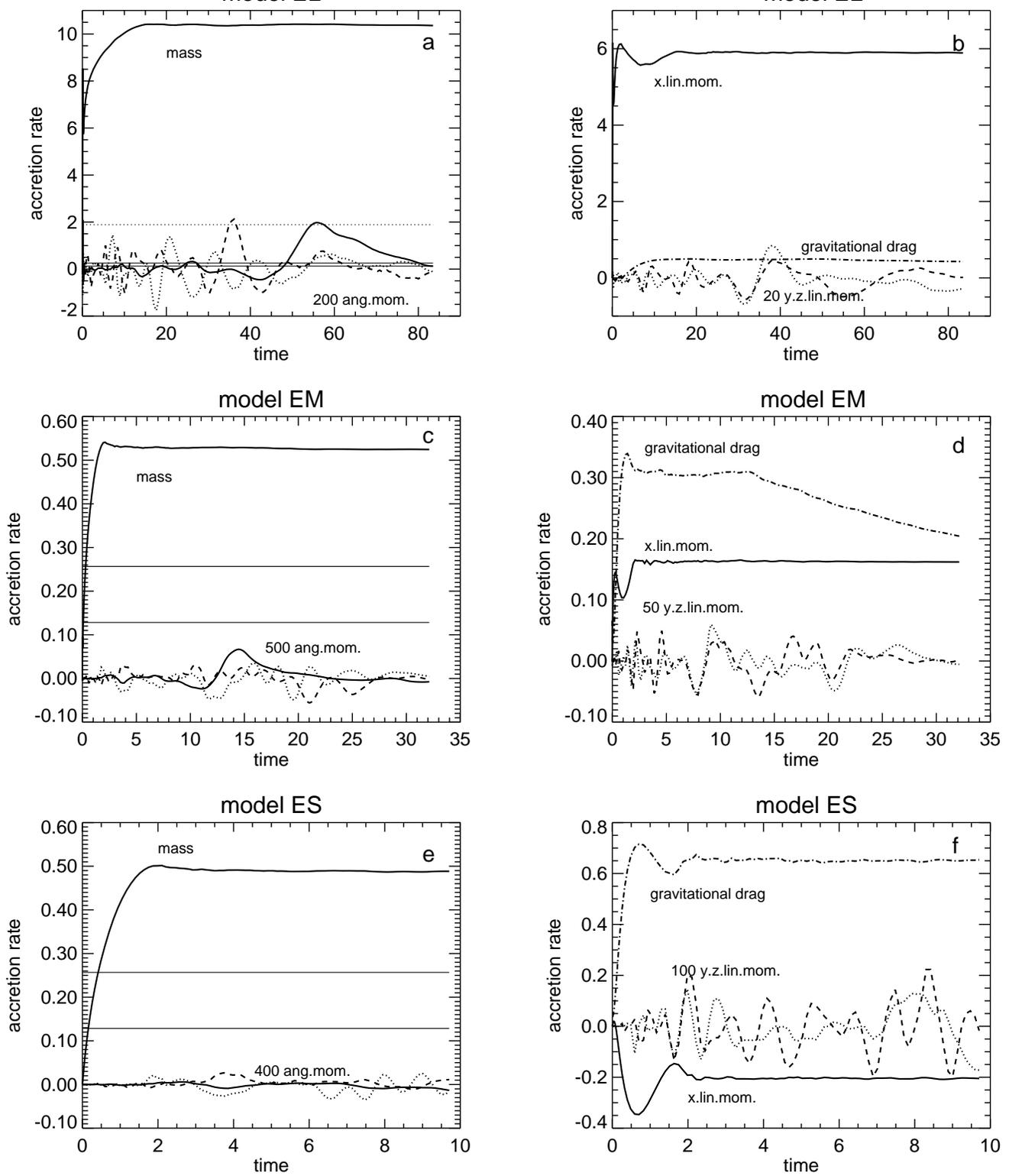

**Fig. 2.** The accretion rates of several quantities are plotted as a function of time for all subsonic models EL, EM and ES. The left panels contain the mass and angular momentum accretion rates. The straight horizontal lines show the analytical mass accretion rates: dotted (panel a) is the Hoyle-Lyttleton rate (Eq. 1 in Paper I), solid are the Bondi-Hoyle approximation formula (Eq. 3 in Paper I) and half that value. The upper solid bold curve represents the numerically calculated mass accretion rate. The lower three curves of the left panels trace the x (solid), y (dashed) and z (dotted) component of the angular momentum accretion rate. The components are multiplied by the factor written down near the curves in order to amplify the deflections of the curves. The accretion rates of the components of the linear momentum (x solid, y dotted and z dashed) are plotted in the right panels. The components are multiplied by the factor written down near the curves in order to amplify the deflections of the curves. Additionally, the gravitational drag (x component) due to the asymmetrical mass distribution upstream and downstream from the accretor is shown as dash-dotted curve.

**Table 1.** Parameters and some computed quantities for all models. $\mathcal{M}_\infty$ is the Mach number of the unperturbed flow, $L$ size of the largest grid, $R$ radius of the accretor, $g$ = number of grid nesting depth levels, $\delta$ = the size of one zone on the finest grid, $\epsilon$ is the softening parameter (zones) for the potential of the accretor (see Paper I), $t_f$ is the total time of the run, $\overline{M}$ is the integral average of the mass accretion rate, $\widehat{M}$ is the maximum mass accretion rate, $S$ is one standard deviation around the mean $\overline{M}$ of the mass accretion rate fluctuations, $\dot{M}_{BH}$ is defined in Eq. 3 of Paper II, $P$ is the approximate period of the accretion rate fluctuations ("vari": no periodicity is visible), $t_u$ is the approximate time at which the flow becomes unstable (the time at which the fluctuations of the ang. mom. accretion rate distincly rise over the background noise; this is the case when a value of 0.2 is reached), $d_{ss}$ shock standoff distance (measured from the center of the accretor; negative:downstream distance of the extrapolated shock lines), $s$ is the entropy (Eq. 4 in Paper II); $l_{rms}$ is the root mean square of the specific angular momentum magnitude; all time units are $R_A/c_\infty$; all models have a specific heat ratio of $\gamma = 1.01$, the initial density distribution is randomly perturbed by 3%, and the number $N$ of zones per grid dimension is 32.

| Model | $\mathcal{M}_\infty$ | $L$ ($R_A$) | $R$ ($R_A$) | $g$ | $\delta$ ($R_A$) | $\epsilon$ | $t_f$ | $\overline{M}$ ($\dot{M}_{BH}$) | $\widehat{M}$ ($\dot{M}_{BH}$) | $S$ ($\dot{M}_{BH}$) | $P$ | $t_u$ | $d_{ss}$ ($R_A$) | $s$ ($\mathcal{R}$) | $l_{rms}$ ($R_A c_\infty$) |
|---|---|---|---|---|---|---|---|---|---|---|---|---|---|---|---|
| ES | 0.6 | 32. | 0.02 | 9 | 1/256 | 3 | 9.71 | 1.90 | 1.91 | 0.01 | $>t_f$ | $>t_f$ | - | 0.26 | 0.001 |
| EM | 0.6 | 32. | 0.1 | 7 | 1/64 | 4 | 32.1 | 2.05 | 2.06 | 0.01 | $>t_f$ | $>t_f$ | - | 0.12 | 0.001 |
| EL | 0.6 | 128. | 1. | 6 | 1/8 | 5 | 83.5 | 40.5 | 40.6 | 0.07 | $>t_f$ | $>t_f$ | - | 0.12 | 0.001 |
| FS | 1.4 | 32. | 0.02 | 9 | 1/256 | 3 | 28.7 | 2.54 | 2.54 | 0.01 | $>t_f$ | $>t_f$ | 0 | 2.98 | 0.001 |
| FM | 1.4 | 32. | 0.1 | 7 | 1/64 | 4 | 29.4 | 2.54 | 2.55 | 0.01 | $>t_f$ | $>t_f$ | 0 | 1.41 | 0.001 |
| FL | 1.4 | 128. | 1. | 6 | 1/8 | 5 | 187. | 5.74 | 5.83 | 0.03 | vari | $>t_f$ | -1.0 | 0.285 | 0.005 |
| GS | 3.0 | 32. | 0.02 | 9 | 1/256 | 3 | 9.55 | 1.20 | 1.36 | 0.08 | vari | 3.5 | 0 | 31.8 | 0.037 |
| GM | 3.0 | 32. | 0.1 | 7 | 1/64 | 4 | 18.5 | 1.18 | 1.34 | 0.07 | vari | 3.5 | 0 | 15.8 | 0.029 |
| GL | 3.0 | 128. | 1. | 6 | 1/8 | 5 | 65.0. | 2.70 | 2.73 | 0.01 | $>t_f$ | $>t_f$ | -1.0 | 0.253 | 0.001 |
| HS | 10. | 32. | 0.02 | 9 | 1/256 | 3 | 3.03 | 0.83 | 1.01 | 0.06 | $\approx 0.2$ | 0.2 | 0.2–0.4 | 170. | 0.09 |
| HM | 10. | 32. | 0.1 | 7 | 1/64 | 4 | 5.87 | 0.82 | 0.89 | 0.02 | $\approx 0.2$ | 0.5 | 0.0 | 110. | 0.10 |
| HL | 10. | 128. | 1. | 6 | 1/8 | 5 | 26.1 | 2.24 | 2.24 | 0.01 | $>t_f$ | $>t_f$ | -2.0 | 0.206 | 0.001 |

### 3.1. Subsonic flow

The distribution of matter for the three subsonic models can be seen in Fig. 1, panels a, c and e. Since the accretor of model EL is large (one accretion radius), the gravity at and beyond the surface is weak. Thus upstream of the accretor matter gets absorbed faster than gravitational focusing can replenish the volume and so a lower density than at infinity is found around $x = -2$. Downstream of the accretor, at $x = +2$ a slight density enhancement is present due to gravitational focusing. For both models EM and ES, in which the accretor is much smaller (than model EL), gravitation can act more strongly and thus concentric density contours form. However, the center of the density contours does not coincide with the center of gravity which also is the center of the vacuum accreting volume. These features have been observed for the models with $\gamma = 4/3$ and $\gamma = 5/3$ (cf. Fig. 1 in Papers III and IV). One very prominent difference between the models with different $\gamma$ is, however, the extent to which the accretor is eccentric relative to the concentric density contours. While this shift is only marginal in the $\gamma = 5/3$ model SS (Paper III, Panel 1d), it is approximately 0.5 $R$ ($R$ being the radius of the accretor) for the $\gamma = 4/3$ model AS (Paper IV, Panel 1e), and for model ES (Fig. 1e) it is 1 $R$. The shift of the center of the contours is downstream relative to the accretor in all cases.

Since matter is more compressible with decreasing values of $\gamma$, the maximum densities reached in the vicinity of the accretor increases when going from model SS over AS to ES. This behaviour can be seen when inspecting the size of some specific contour, e.g. $\lg \rho = 1$, in the plots of these models.

The accretion rates of several quantities for the subsonic models EL, EM and ES are shown in Fig. 2. A steady state is reached in all three cases with only small fluctuations in the rate of accreted linear and angular momentum. These are due to the initially perturbed medium and they die down after the inflowing unperturbed wind flows through the whole grid. A comparison of the accretion rates of these new models EL, EM, ES with models SL, SM, SS, AL, AM, AS from Papers III and IV reveal no striking differences. Only the sign of the accreted x-component of the linear momentum is clearly negative in model ES (Panel 2f) while it is clearly positive in model SS (Paper III, Fig. 2, bottom right). A positive sign indicates that more momentum is accreted from the forward (upwind) side of the accretor than from the back (downwind).

Although it might seem that model ES is unstable, the fluctuations visible in Panels 2e and f are due to the initial conditions. It takes the flow roughly 27 time units (half the size of the grid divided by the flow speed) to cross the grid from the boundary to the accretor and thus to fill this area with 'fresh' unperturbed material.

Since the models FL, GL and HL, simulating large accretors with size 1.0 $R_A$, all show similar features, we will discuss these models together in this section. All these models reach a steady state in which the mass accretion rate is practically constant and the linear and angular momentum accretion rates are negligible, except, of course, for the x-component of the linear momentum (cf. Panels a and b in Figs. 3, 5 and 7). The value of the momentum's x-component is large and positive, indicating a large amount of momentum being accreted directly geometrically from the upstream flow. Model FL, however, does show fluctuating components of linear and angular momentum accretion rates, albeit, at a very low level. The time it takes the initially perturbed density to cross half the grid is about 50 time units for model FL, so these fluctuations cannot be due to the initial perturbations. The density distribution and shock position at large distances (5 $R_A$ or more) from the accretor are also representative for what the distribution looks like for the models with smaller accretor. They can be seen in Panels 1b, and 4a and b. Obviously, as the Mach number increases the opening angle of the shock cone decreases (cf. Sect. 4.4). In all cases the shock is a static tail shock in which the flow downstream is directed away from the accretor.

### 3.3. Mildly supersonic flow

Contrary to the subsonic models, a shock front develops when the relative velocity between accretor and medium is supersonic: contour plots of the density of the mildly supersonic models FL, FM and FS can be seen in Fig. 1. In all these models the accretor moves at a velocity of Mach 1.4 relative to the surrounding medium. Note the similarity of the contours of panels d and f in Fig. 1, manifest in the shape of the contours and the position and strength of the shock at some fixed distance. This indicates that the density distribution is not very sensitive to the size of the accretor for these models and to the differing mass accretion rates (see Table 1 and the following paragraphs) that they generate. It is due to the fact that the sonic distance is expected to be larger than the size of the accretor, as was mentioned in Section 2.2. Downstream of the accretor in models FM and FS (panels 1d and f) a very spherically symmetric flow develops within the shock cone. Contrary to the models BM, BS, KL and KS (in Papers III and IV, they have larger values of $\gamma$) the shock front in the models with $\gamma = 1.01$ (panels 1d and f) does not clear the surface of the accretor during the whole simulation. The fluctuations eventually die down and a steady state is reached.

The large scale (around 10 $R_A$) density distribution and flow of model FL look very similar to those of the models with smaller accretor, while the flow at a distance of a few accretor radii from the surface of the accretor is somewhat different (see Fig. 1): Although the shock front in all three models ends at the surface and is a tail shock, practically no radial inflow is visible downstream for model FL (except within 0.3 accretor radii distance from the surface).

the grid is about 10 time units for models FM and FS. Figure 3 shows the runs of the accretion rates and of the gravitational drag as a function of time. The plot shows that these models converge toward a steady state after the initial transients have calmed down (the initial perturbations were advected off the grid).

Also in the supersonic models FS and FM the x-component of the accreted linear momentum has a negative sign, indicating that more momentum is accreted from the back (downwind) side of the accretor than from the upwind side. Thus the accretion of linear momentum in these cases would not act as a drag but as an accelerating force! However, since the gravitational force always acts as drag and is stronger than the linear momentum accretion drag, the net force is decelerating.

### 3.4. Moderately supersonic accretors

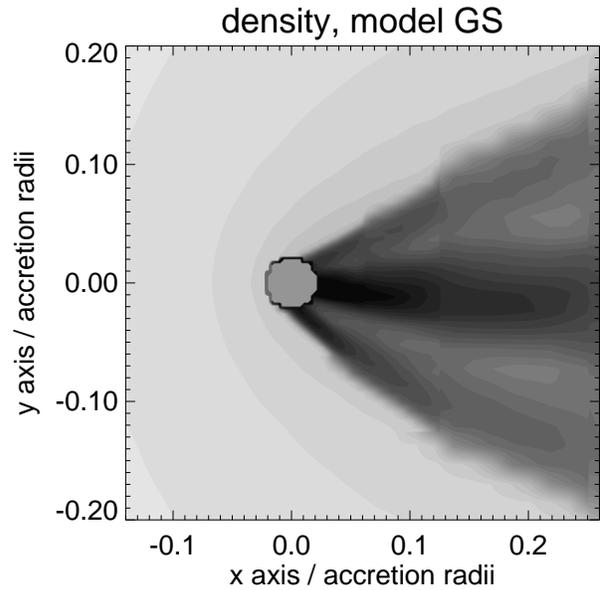

**Fig. 6.** Gray scale plots showing the density distribution of model GS in a plane containing the center of the accretor. Darker shades represent higher densities. The time of the snapshot is the same as in Fig. 4e.

Compared with all other simulations presented in this work, the moderately supersonic models GM and GS exhibit the most unstable flow and the largest fluctuations, e.g. the mass accretion rate varies by roughly 7% (cf. Table 1). This is, however, small compared to the models with larger $\gamma$ that produce mass accretion rate fluctuations of 14% (for $\gamma = 4/3$) and of 23% (for $\gamma = 5/3$). The comparison of different rates is done in more detail in Sect. 4.1.

The contour plots of the density distribution (Fig. 4, panels c and e) show a striking difference to the equivalent plots of the models at different $\gamma$ (cf. Paper IV, Figs. 4c and 4e; Paper II, Figs. 6 and 11): the shock cone ends at the surface, i.e. it density distribution, one notices that the shock never clears the surface. This trend had already been apparent when comparing the results of the simulations for $\gamma = 5/3$ to $\gamma = 4/3$ (in Paper IV) – the shock front was closer to the accretor surface and touched the surface more often – and is now (with $\gamma = 1.01$) even more extreme. So although the strength of the fluctuations is reduced in these models in which the shock ends at the accretor surface, nevertheless, the flow is still unstable resulting in non-radial accretion flow. Lower density pockets, vortices etc. can be discerned.

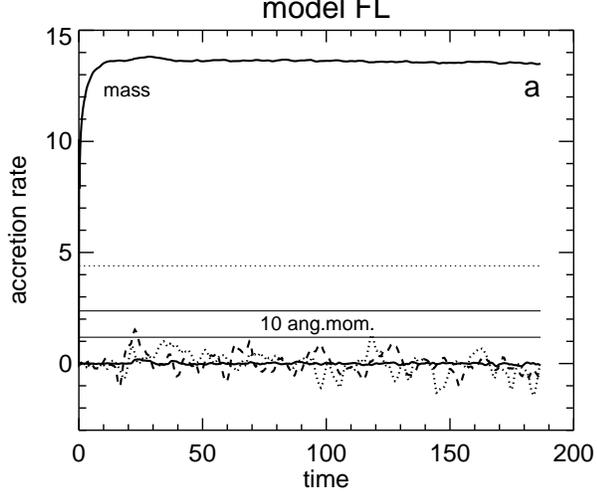
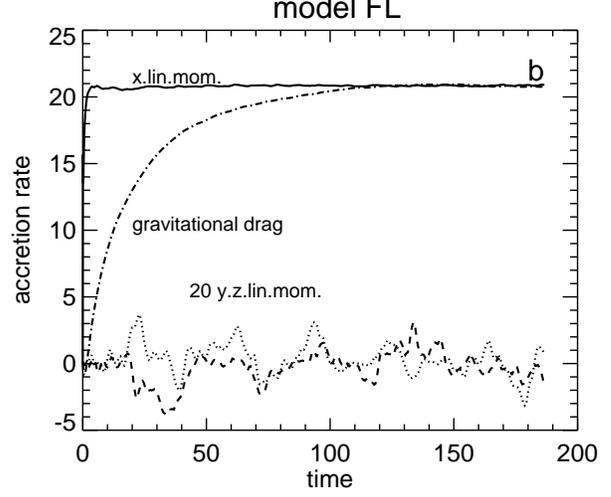
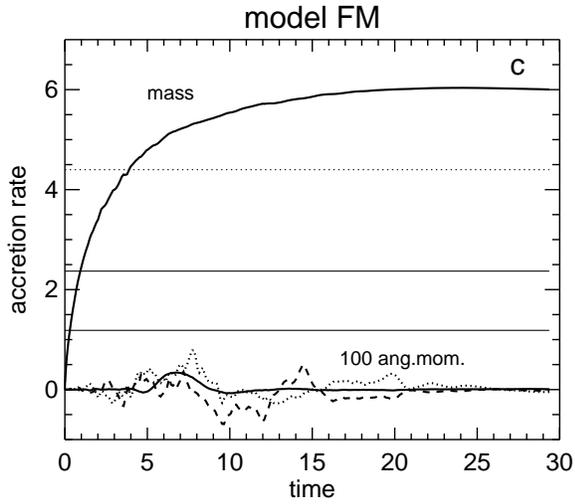
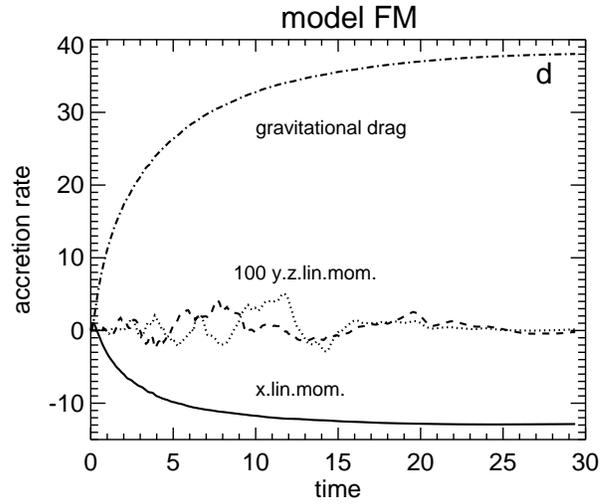
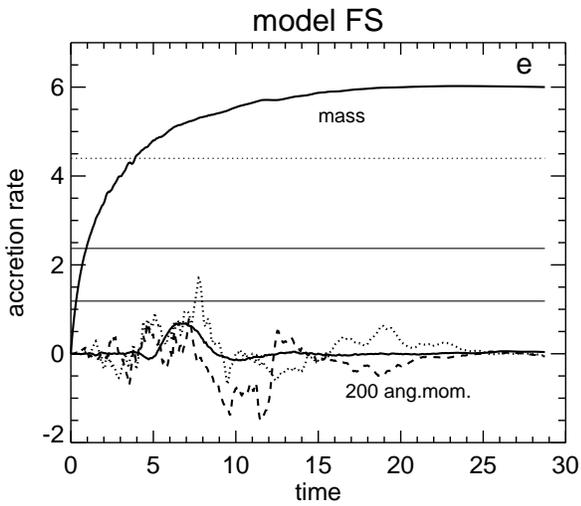
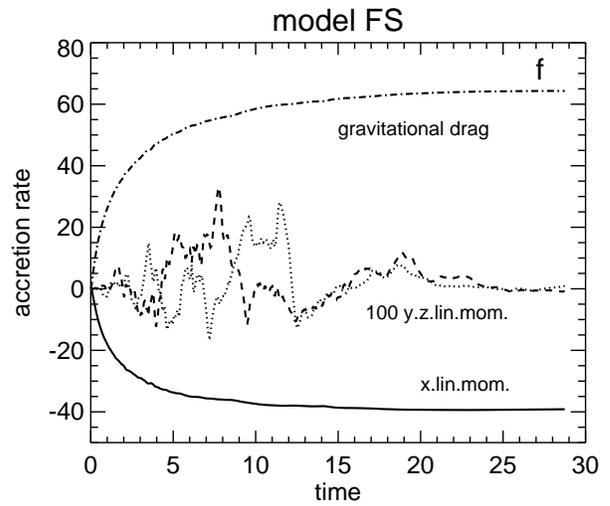

**Fig. 3.** The accretion rates of several quantities are plotted as a function of time for all mildly supersonic models FL, FM and FS. The left panels contain the mass and angular momentum accretion rates, the right panels the linear momentum accretion rates and the gravitational drag. The components are multiplied by the factor written down near the curves in order to amplify the deflections of the curves. See also the caption of Fig. 2

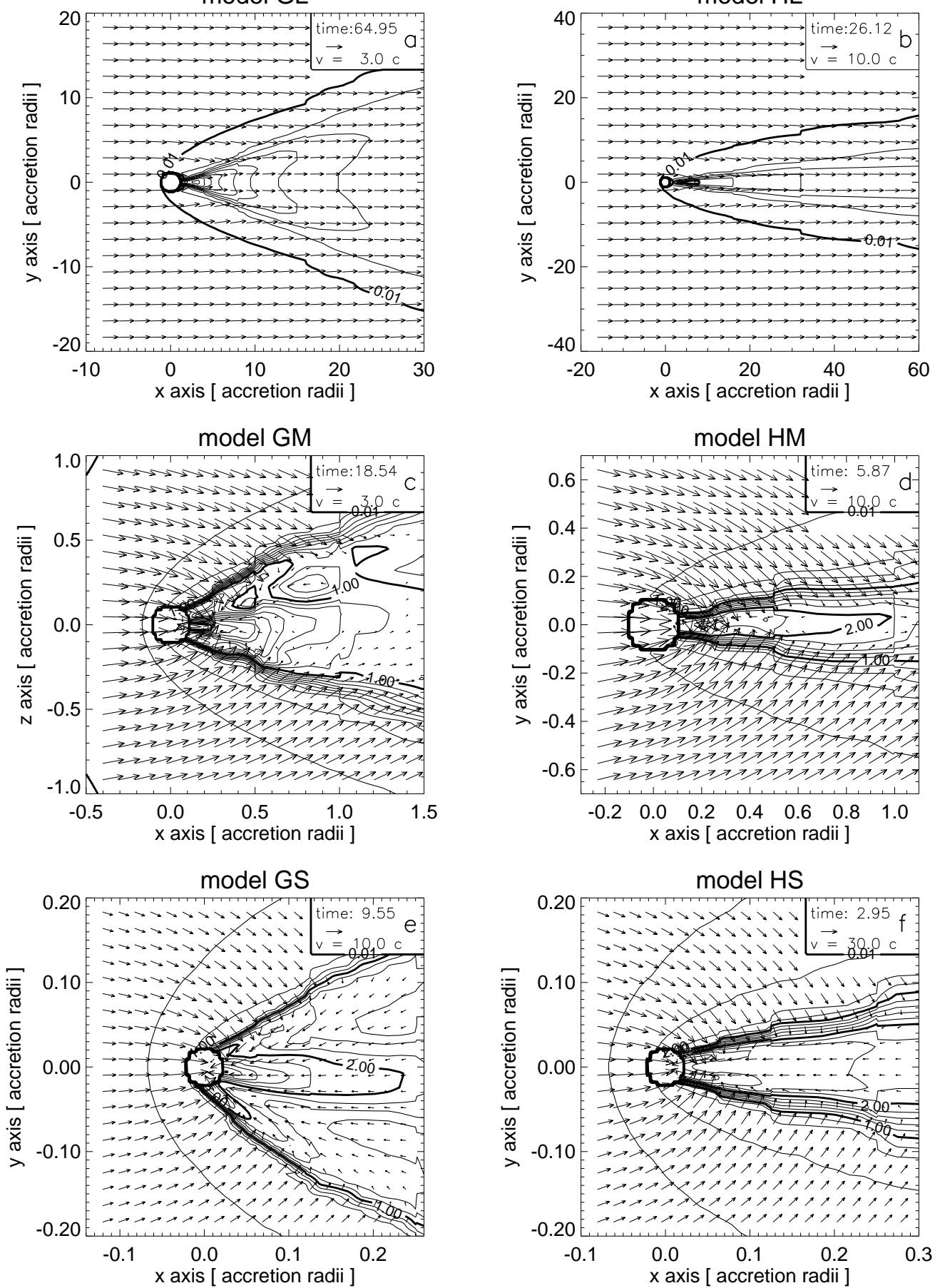

**Fig. 4.** Contour plots showing snapshots of the density together with the flow pattern in a plane containing the center of the accretor for all moderately supersonic models GL (panel a), GM (panel c), GS (panel e) and highly supersonic models HL (panel b), HM (panel d), HS (panel f). The contour lines are spaced logarithmically in intervals of 0.1 dex for models GL, HL and GM, and 0.2 dex for the other models. The bold contour levels are labeled with their respective values. The time of the snapshot together with the velocity scale is given in the legend in the upper right hand corner of each three panels.

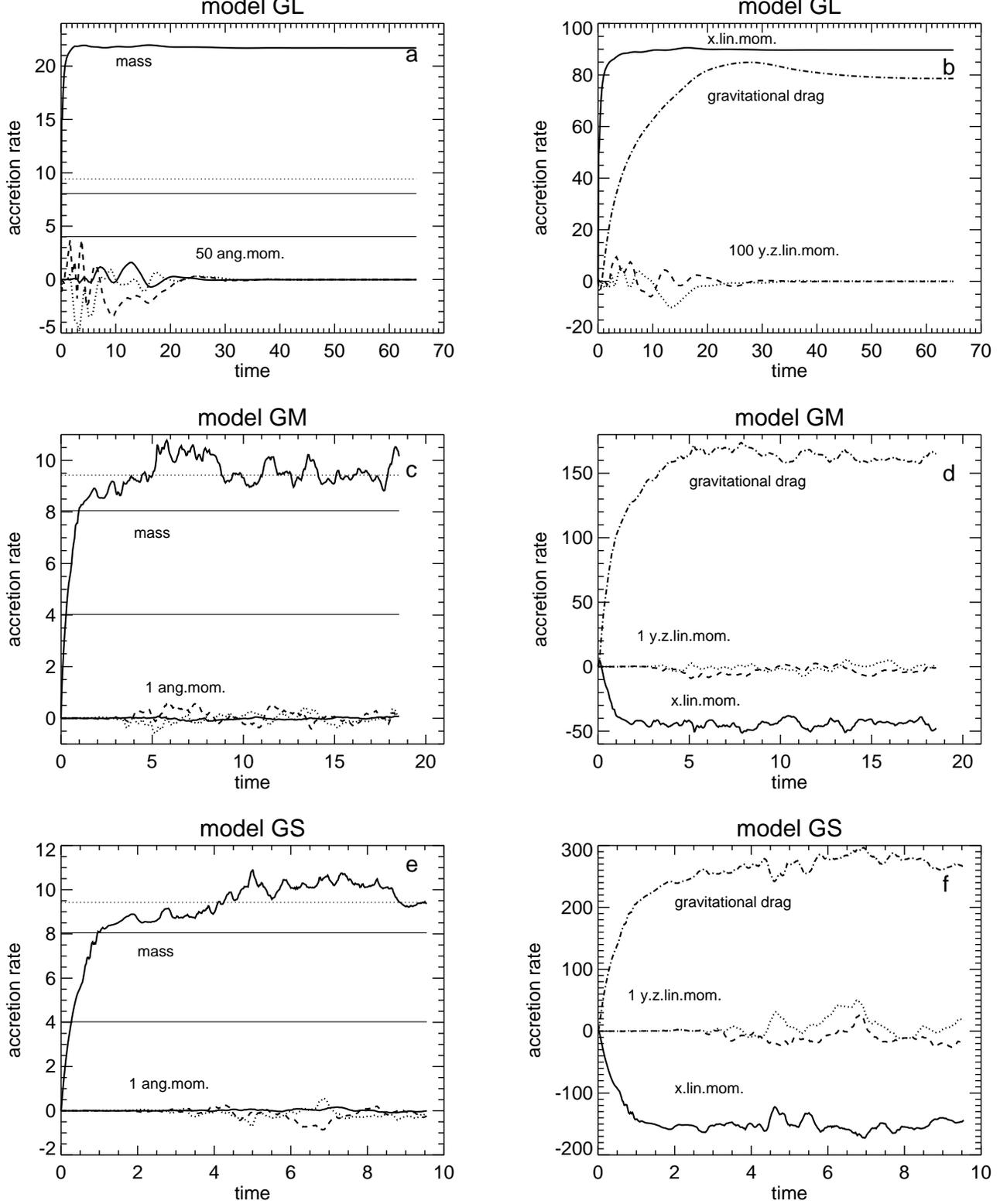

**Fig. 5.** The accretion rates of several quantities are plotted as a function of time for all moderately supersonic models GL, GM and GS. The left panels contain the mass and angular momentum accretion rates, the right panels the linear momentum accretion rates and the gravitational drag. The components are multiplied by the factor written down near the curves in order to amplify the deflections of the curves. See also the caption of Fig. 2

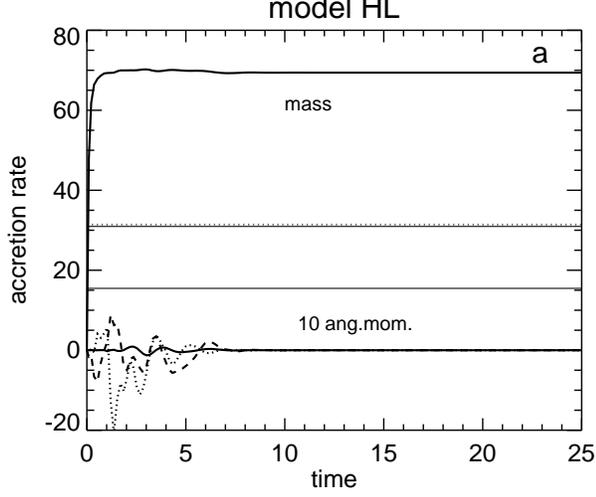
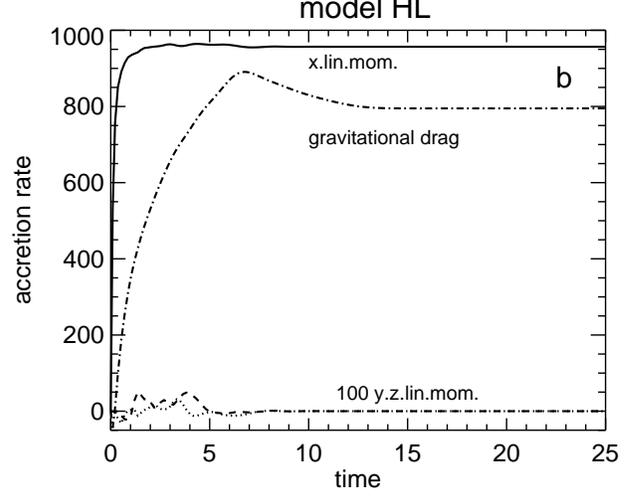
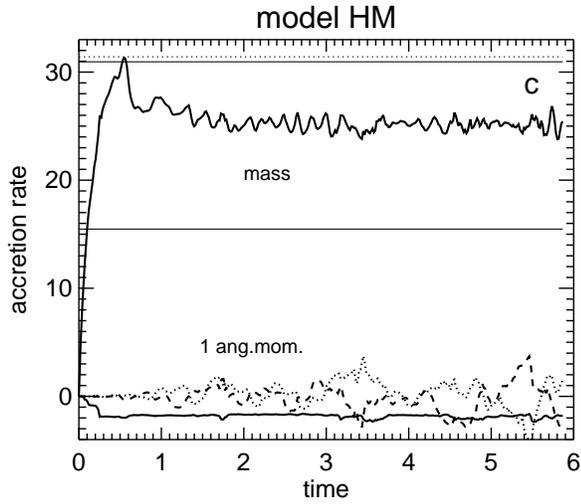
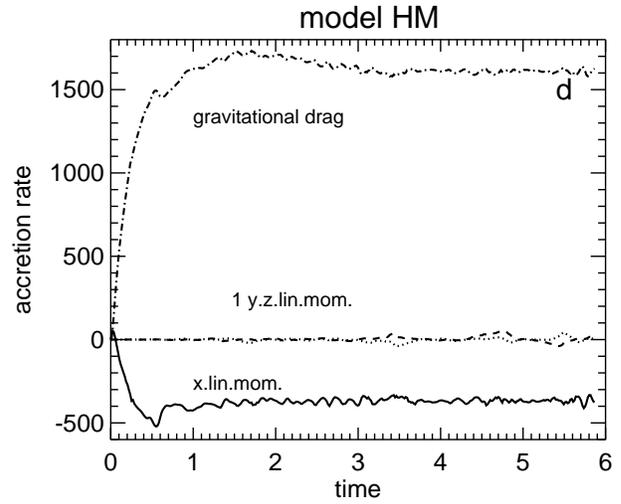
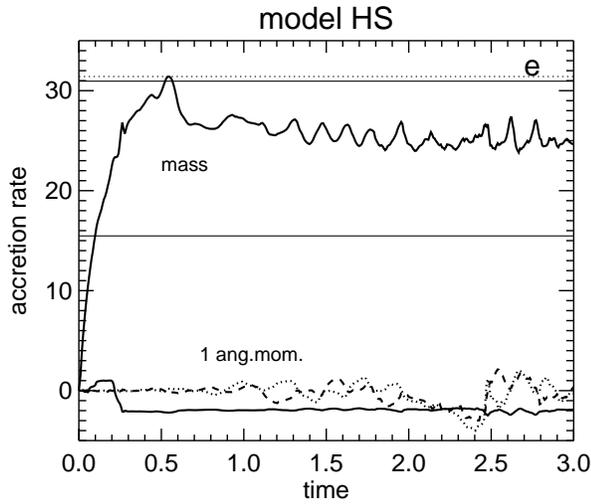
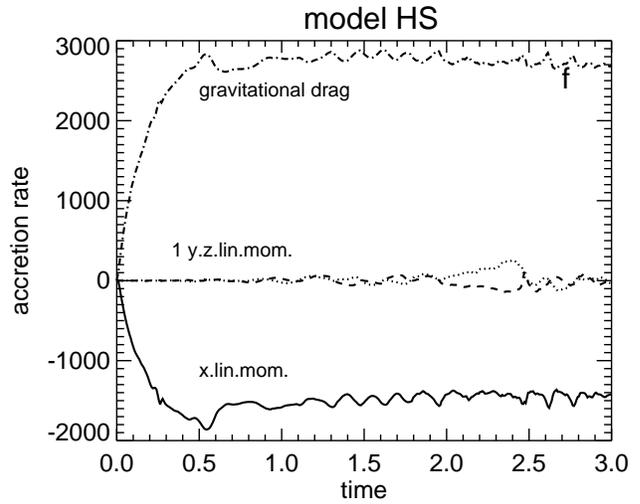

**Fig. 7.** The accretion rates of several quantities are plotted as a function of time for all highly supersonic models HL, HM and HS. The left panels contain the mass and angular momentum accretion rates, the right panels the linear momentum accretion rates and the gravitational drag. The components are multiplied by the factor written down near the curves in order to amplify the deflections of the curves. See also the caption of Fig. 2

Two regions with a local density maximum are visible in Panel 4e and perhaps better visible in Fig. 6 : (a) directly downstream in the wake of the accretor an accretion line forms and (b) directly along the shock cone after the infalling material has crossed the shock. It is similar to the mushroom shaped density distribution that has been described in Paper II, and that has preceded the unstable flow. In the new models presented in this paper the structure is fixed to the surface of the accretor, while for the $\gamma = 5/3$ flows the separate density maximum appeared when the shock was clearing the surface of the accretor.

When inspecting the time evolution of the fluctuating accretion rates of mass, angular and linear momentum (Fig. 5, Panels c–f) one sees the unstable flow reflected in the fluctuating accretion rates. Obvious periodicities are not visible.

### 3.5. Highly supersonic accretors

The highly supersonic models HM and HS, too, do not converge to a quiescent steady state but show an unstable fluctuating flow pattern. However, the fluctuations are less pronounced than in models GM and GS. In Panels 4d and f, snapshots of the typical density distribution have been plotted for the two models HM and HS, respectively. One cannot help but notice the very regular flow pattern and shock cone structure: the position of the shock cone hardly changes (in time) and the shape is practically straight. Thus Panels d and f (of Fig. 4) just seem to be an enlargement of Panel b. Note, however, (a) the shock strength (more contour lines over a shorter distance) increases with decreasing size of the accretor and (b) there is a stagnation point at a position of about $x = 0.8$ (Panel 4d). Further out than this distance the flow escapes from the accretor (Panel b), while within this distance matter is accreted in a stream collimated toward the accretor (Panel f). This flow structure is very much different from the very complicated state shown in Fig. 8e in Paper IV (which is the equivalent model for $\gamma = 4/3$).

The temporal evolution of the values of some accreted quantities can be seen in Fig. 7. They show the slightly unstable accretion of mass and momenta. Contrary to the models GM and GS, the models HM and HS exhibit a periodicity with a period of approximately 0.2 time units, which can be seen from the Fourier transform of the mass accretion rates (Fig. 8).

## 4. Analysis of Results

### 4.1. Mass accretion rate

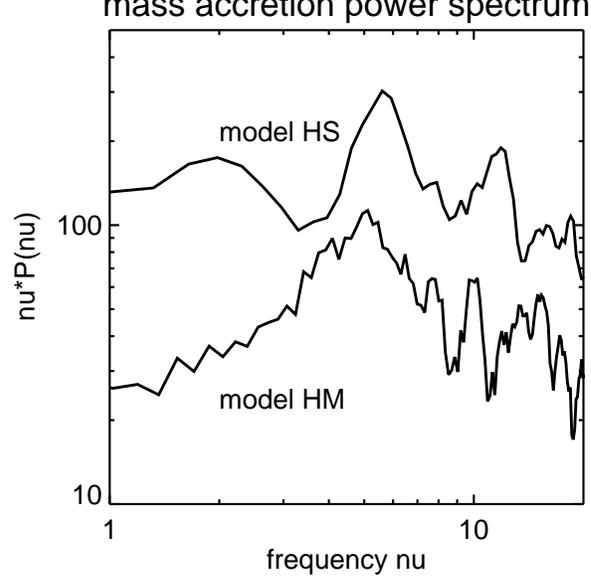

**Fig. 8.** The power spectrum $\nu P(\nu)$ (in arbitrary units) of the mass accretion rate for models HM and HS.

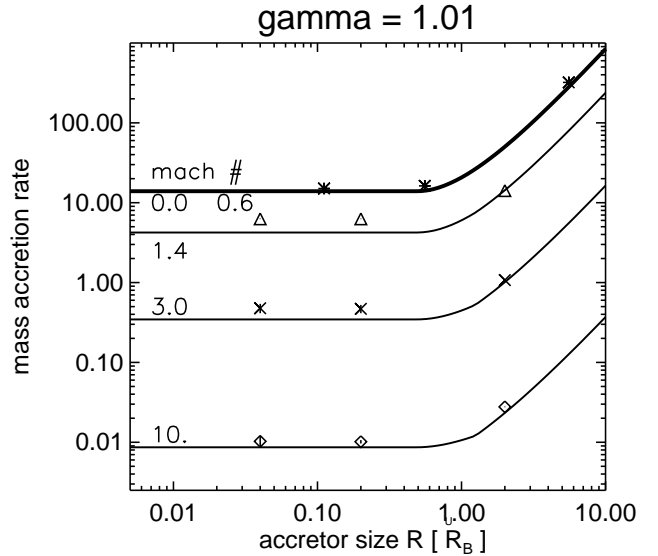

**Fig. 9.** Mass accretion rates (units: $4\pi R_{\rm B}^2 c_\infty \rho_\infty$) as a function of accretor size (units: $\breve{R}_{\rm B}$ (Eq. 7 in Paper III), *not* accretion radii). The Mach number is an additional parameter and is specified by the numbers to the right of the y-axis. The topmost bold curve corresponds to a Mach number of zero, i.e. a stationary accretor. The other lines belong to Mach numbers of 0.6, 1.4, 3.0 and 10., whereby the curve of Mach 0.6 completely overlaps with the curve belonging Mach 0. The numerical results to the corresponding Mach numbers are plotted with different symbols: the stars ($\ast$), triangles ($\triangle$) crosses ($\times$) and diamonds ($\diamond$) are the results of models with $\mathcal{M}_\infty$=0.6, 1.4, 3.0 and 10, respectively. The "error bars" extending from the symbols indicate two standard deviations from the mean ($S$ in Table 1), for models in which the mass accretion rate fluctuates.

Fig. 9. Note that in this figure the unit of length (for the accretor size) is $\check{R}_B$ (Eq. 7 in Paper III), *not* $R_A$. In Paper III we had described an interpolation formula that had well fit (in the spirit of Bondi, 1952) the mass accretion rates found in Papers I, II and III. Since all those models were done using a specific heat ratio of 5/3, the formula is adapted only to that one $\gamma$. In Paper IV the numeric mass accretion rates were compared to the values obtained with this formula and only a slight difference was found. In Fig. 9 we include the mass accretion rate that follows from our interpolation formula for $\gamma = 1.01$ as solid curves.

Two points seem noteworthy. (a) As has been mentioned in Sect. 2.2 the sonic distance for a gas with $\gamma = 1.01$ is $d_s = 0.49 R_B$. So just as was the case in Paper IV, the curves run horizontally for all accretors that have a radius smaller than the sonic distance. Within the sonic radius the flow is supersonic and changes of the size of accretors smaller than the sonic distance should not influence the accretion flow. Within numerical accuracy, the mass accretion rates derived numerically are equal for those models in which the accretor size is smaller than the sonic distance, e.g. the mass accretion rate for models HM and HS are 0.82 and 0.83, respectively (cf. Table 1). (b) The interpolation formula does not fit the mass accretion rates for all supersonic models too well: the numerical values lie above the interpolation curves by up to 60%. Especially the models with smaller Mach number deviate significantly from the formulae.

The flow patterns are always less violent in the $\gamma = 1.01$ models compared to the $\gamma = 5/3$ or $\gamma = 4/3$ models: Fig. 10 shows the relative mass fluctuations, i.e. the standard deviation divided by the average mass accretion rate $\overline{M}/S$ (cf. Table 1), for all pertinent models. One notices that in all cases the largest bold symbols lie below the less large and bold symbols of same shape and same Mach number. This result extends the trend described in Paper IV refering to $\gamma = 4/3$ compared to $\gamma = 5/3$ models. Again, the fluctuations of model HM are much smaller than those of model GM, just as had been the case with models DM and CM in Paper IV. It is due to the fact that in the models with high Mach number the shock hardly ever clears the surface of the accretor and when it does, it does not venture very far. Thus, the part of the accreted quantities that are accreted directly from the incoming stream (i.e. without having to cross the shock cone) remain constant and do not participate in the unstable accretion flow.

The maximum density that is reached on the numerical grid is shown in Fig. 11. In the same way as was described in Paper IV (and also in Shima et al., 1985) this density increases in models with (1) higher Mach numbers, (2) smaller accretors and (3) lower $\gamma$-values. Thus, these three points are only confirmed and extended in their trend by the new $\gamma = 1.01$ models.

### 4.2. Angular momentum accretion rate

For many applications the magnitude of the accreted specific angular momentum is of interest. In Fig. 12 we plot for several models the ratio of temporal rms average of the specific angular momentum $l_{rms}$ over Mach number $\mathcal{M}_\infty$ as a function of the accretor size. The values of $l_{rms}$ and $\mathcal{M}_\infty$ can be found in Table 1. Additionally, at the top left of the plot, the diagonal line shows what values are expected analytically if one assumes that a vortex flows with Kepler velocity $V$ around the accretor, just above the surface, i.e. $l_{rms}$ is given by $l_s = RV = \sqrt{R}\mathcal{M}_\infty$. This plot (Fig. 12) covers approximately the same range as Fig. 14 in Paper IV. Only those models that exhibit unstable, active flow patterns are shown.

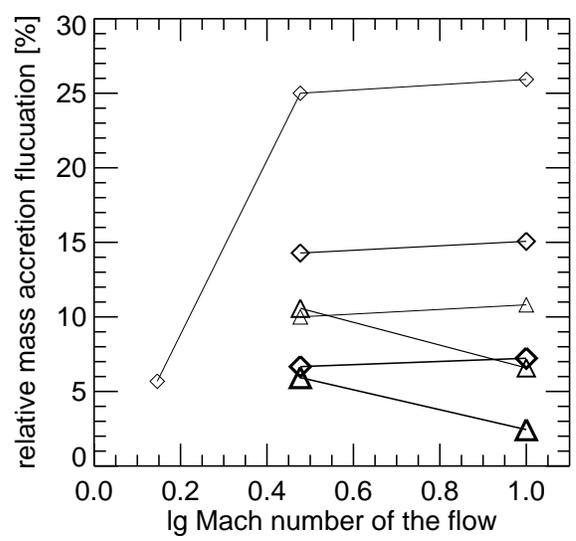

**Fig. 10.** The relative mass fluctuations, i.e. the standard deviation $S$ divided by the average mass accretion rate $\overline{M}$ (cf. Table 1), is shown as a function of the Mach number of the flow. Diamonds ($\diamond$) denote models in which the accretor has a radius of 0.02 $R_A$, triangles ($\triangle$) models with 0.1 $R_A$. The largest bold symbols belong to models with $\gamma = 1.01$, while the smaller symbols belong to models with $\gamma = 4/3$, and the smallest thin symbols are $\gamma = 5/3$. The lines connecting the symbols are for reference only.

Note that the angular momentum accreted in the models presented in this paper ($\gamma = 1.01$) lie well below the values produced by the $\gamma = 5/3$ models of Papers II and III, and by the $\gamma = 4/3$ models of Paper IV. This confirms the fact that has already been mentioned: the fluctuations of the $\gamma = 1.01$ models are smaller than of the models with larger $\gamma$.

### 4.3. Linear momentum accretion rate and drag forces

Figures 13 and 14 are a collection of drag force values for all models that we have calculated upto now, from the series of Papers II to IV and the new results of the present paper. The amount of linear momentum accreted in all models is shown in Fig. 13. This linear momentum is what induces a drag force that can be called hydrodynamical drag. In the actual physical situation the asymmetry of the mass density distribution upstream and downstream from the accretor produces an additional drag force, that is called gravitational drag. This gravitational drag depends, however, logarithmically on the geometric extent of the surrounding medium through which the accretor moves (Chandrasekhar dynamical friction equations, e.g. in Binney & Tremaine, 1987; Chandrasekhar, 1943). For this reason the units of the gravitational drag values shown in Fig. 14 include the term $\ln(L/R_A)$, in which $L$ is the size of the largest grid. One has to remember to remultiply this logarithmic term using the correct size of the medium when applying the plotted values to physical situations.

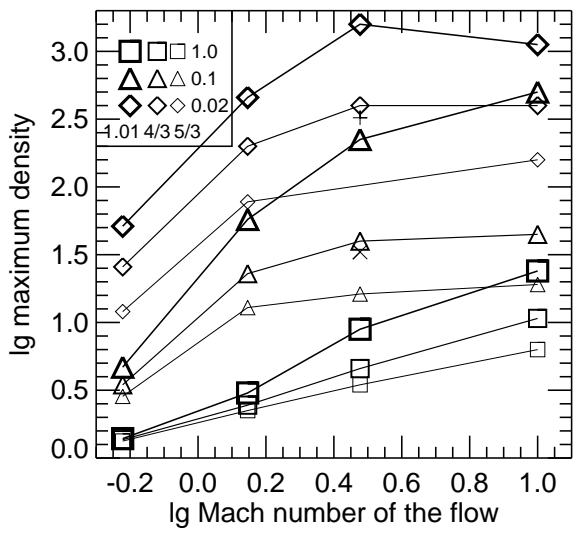
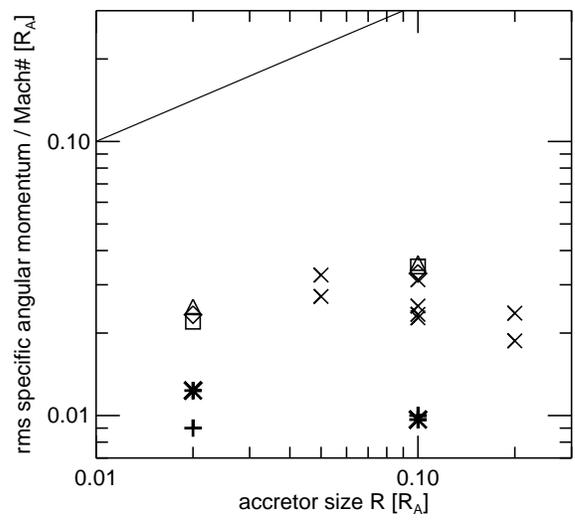

**Fig. 11.** The maximum density that is reached on the grid is shown as a function of the Mach number of the flow. Diamonds ($\diamond$) denote models in which the accretor has a radius of $0.02\ R_A$, triangles ($\triangle$) models with $0.1\ R_A$, squares ($\square$) models with $1.0 R_A$, while the cross ($\times$) is model S6 with $0.2 R_A$ and the plus (+) is model T10 with $0.01 R_A$. The largest bold symbols belong to models with $\gamma = 1.01$, while the smaller symbols belong to models with $\gamma = 4/3$, and the smallest thin symbols are $\gamma = 5/3$. The legend in the top left corner summarises how the nine symbols depend on the three values of $\gamma$ and on the three values of the accretor radius. The lines connecting the symbols are for reference only.

**Fig. 12.** As a function of accretor size the ratio of temporal rms average of the specific angular momentum $l_{\rm rms}$ over Mach number $\mathcal{M}_\infty$ is plotted for all models: squares ($\square$) for models FS and FM from Paper III, crosses ($\times$) for the models from Paper II, diamonds ($\diamond$) for models CM and CS from Paper IV, triangles ($\triangle$) for models DM and DS from Paper IV, large bold stars ($*$) for models GM and GS and finally large bold plus-signs (+) for models HM and HS. The diagonal straight line at the top left shows what values $l_s$ (see text) are obtained if the flow would rotate with Kepler velocity along the accretor surface.

A hydrodynamic *drag* force, contrary to an accelerating force, is actually only felt by the models simulating large accretors, with radii of one $R_A$. In these cases, the geometric accretion of momentum from matter flowing directly from the upstream direction into the accretor (possibly slightly focussed by gravitation) dominates any other accretion from matter downstream. All subsonic models ($\mathcal{M}_\infty = 0.6$) produce hydrodynamic drag forces that are larger (more positive) than the corresponding supersonic models with other parameters kept equal. Thus the shock cone in the supersonic models seems to collimate the linear momentum accretion in the negative x-direction more strongly than the basically radial spherically symmetric but eccentric flow present in the subsonic models. Although the velocities are roughly of similar magnitude close to the accretor, independent of whether the matter has passed the shock or not, the density is larger within the shock cone, so the momentum accreted from the shock cone is larger than the momentum accreted from the upstream direction. For this reason the momentum accreted by the supersonic models has a larger share of negative momentum. Thus most models of interest (supersonic and not large) actually experience an accelerating momentum accretion force. Only the smallest subsonic models (ES and EM) have a slightly negative accreted momentum, too, generated by the very eccentric density contours (cf. Fig. 1), that produce a high density just downstream of the accretor. The gravitational force always acts as drag (decelerating) and is stronger than the momentum accretion drag in all our models, so the net force is decelerating. However, it seems in principle possible to construct a situation such that the gravitational drag is smaller than the magnitude of the hydrodynamic drag. The latter can be negative, so a net acceleration would result. Of course, energy and momentum conservation dictates, that this situation cannot be a stationary one.

Another feature visible in Fig. 13 is that with decreasing $\gamma$ the accreted momentum becomes more negative for models with accretor sizes $0.02\ R_A$ and $0.1\ R_A$. This, too, can be traced to the more regular and better collimated flow in the models with smaller $\gamma$, which can directly be seen when comparing the contour plots, e.g. Fig. 4e and f with Fig. 6e and Fig. 8e from Paper IV.

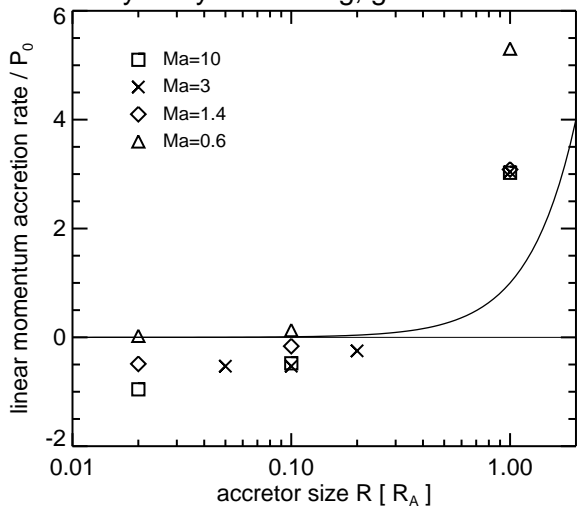
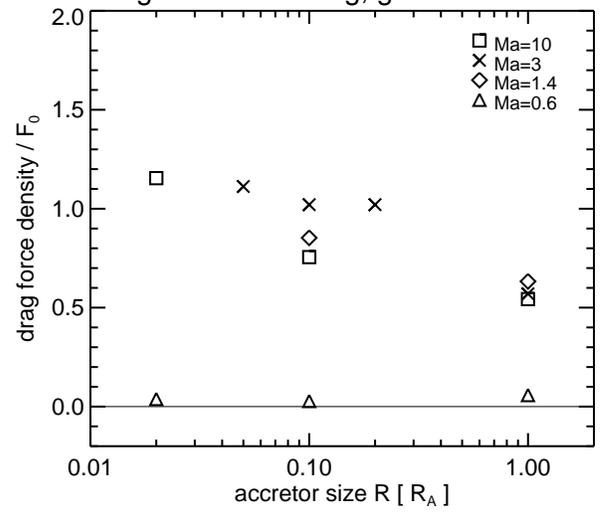
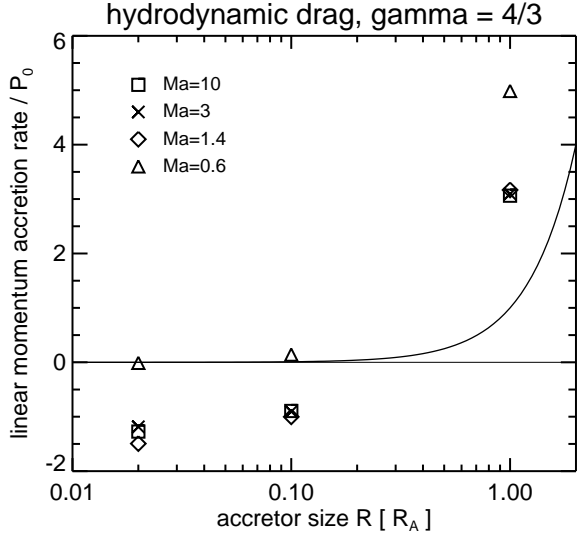
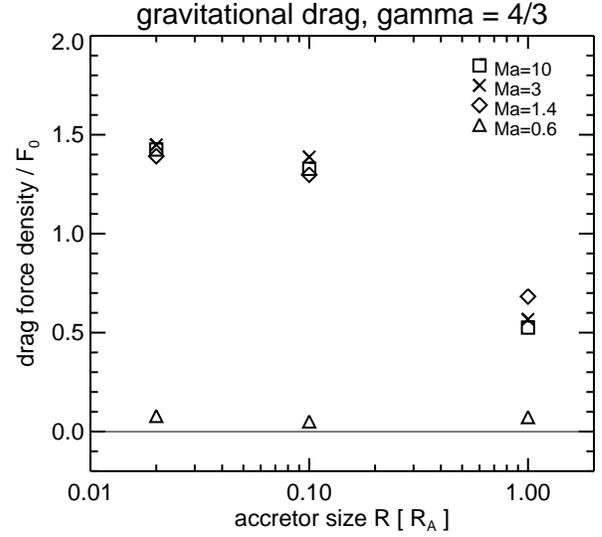
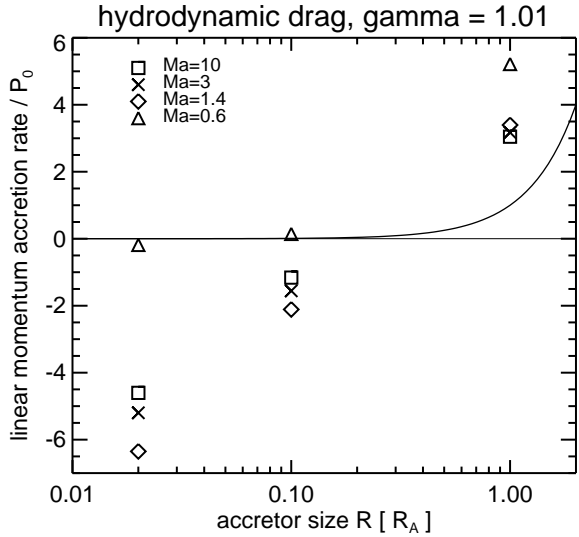
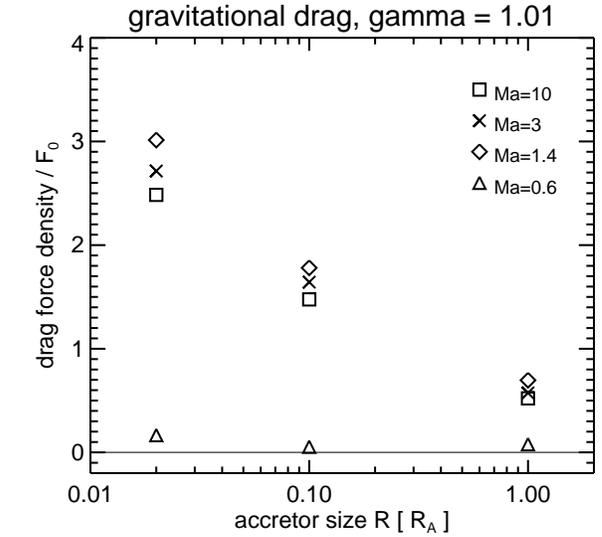

**Fig. 13.** The linear momentum accretion rate is plotted as a function of accretor size for all models: the $\gamma = 5/3$ values are taken from Papers II and III, the $\gamma = 4/3$ values from Paper IV and the $\gamma = 1.01$ values are shown in Figs. 2, 3, 5, and 7. The bold curve over the horizontal straight line (separating positive and negative values) represents the amount of momentum that would be accreted purely through the geometric cross section (i.e. $\pi R^2 \rho v_\infty^2$). The units are $P_0 = \pi R_A \rho v_\infty^2$; negative values indicate an acceleration, positive values a drag.

**Fig. 14.** The gravitational drag force density is plotted as a function of accretor size for all models: the $\gamma = 5/3$ values are taken from Papers II and III, the $\gamma = 4/3$ values from Paper IV and the $\gamma = 1.01$ values are shown in Figs. 2, 3, 5, and 7. The units are $F_0 = P_0 \ln(L/R_A)$, with $P_0$ defined in the caption of Fig. 13, and $L$ is the size of the largest grid (cf. Table 1). The zero line is plotted for reference only.

or negligible (within numeric accuracy) gravitational drag. Just as with the hydrodynamic drag (Fig. 13) the spread of values due to different Mach numbers of the flow is of the same order of magnitude as the spread due to different accretor sizes. A clear trend can be seen, that (a) with decreasing accretor sizes (b) for smaller Mach numbers and (c) for smaller $\gamma$, the gravitational drag force increses. The softer equation of state in models with smaller values of $\gamma$ yields larger accretion cone densities and thus stronger gravitational forces. Although flows with higher Mach numbers produce stronger shocks and thus higher densities within the shock cone, too, the shock openening angle in these fast flows is smaller. Thus the gain in gravitational drag due to the higher density seems to be offset by the smaller shock cone. The density increases close to the accretor, so it is not surprising that models with smaller accretor sizes show a larger gravitational drag: the front-back asymmetry is larger. Note the big difference in magnitude of the gravitational drag when comparing the subsonic models (Mach 0.6) with the mildly supersonic models (Mach 1.4): the drag increases by an order of magnitude. A further increase of the Mach number from 1.6 to 10, on the other hand, *reduces* by roughly 30% the gravitational drag (when assuming a fixed accretor size). Thus the presence of the shock cone is crucial for a non-negligible gravitational drag force.

*4.4. Comparison of shock opening angles*

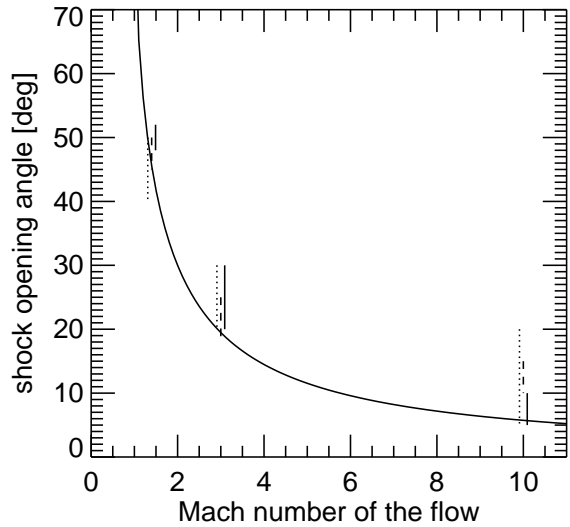

**Fig. 15.** The shock opening angles for various models are plotted versus speed of the flow. The vertical lines show the numerical models, solid represents the $\gamma = 1.01$ models, dashed the $\gamma = 4/3$ models and dotted the $\gamma = 5/3$ models. The vertical lines have slightly been shifted horizontally to facilitate the distinction of the lines. The curve decreasing shows the relation $\Theta = \arcsin(1/\mathcal{M})$.

per IV), the shock opening angle ($\theta$, angle of the shock front to the direction of the unperturbed flow) seen in the simulations tends to be larger than the values predicted analytically, $\Theta = \arcsin(1/\mathcal{M})$, for large distances from the accretor. A rough estimate of the shock opening angle can be obtained from the contour plots and the values are plotted in Fig. 15 by vertical bars, the range of which span the different angles of the inner and outer shock cone edges, and of the different models, at large distances from the accretor.

Contrary to the results presented in Paper IV for $\gamma = 4/3$, the deviation of the numerically obtained angles from the analytic curve for models with $\gamma = 1.01$ is about equal for all Mach numbers. The suggestion by other authors (e.g. Livio et al., 1991) that it is the pressure within the shock cone that enlarges and expands the boundaries of the cone (thus increasing the opening angle) seems to be corroborated: for the models with a large Mach number ($\mathcal{M}_\infty = 10$) the shock angle decreases with decreasing $\gamma$. Smaller $\gamma$ imply a softer equation of state which potentially produces a smaller pressure. That this effect is strongest at large Mach numbers is due to the strong shock and thus high density and pressure generated in the wake.

## 5. Conclusions

We draw the following conclusions from the results presented here:

1. For $\gamma = 1.01$ the flow (and thus the accretion rate of all variables) exhibits active unstable phases only for models in which the flow speed is large enough and the accretor size small enough: all models with Mach numbers of 0.6 or 1.4 tend toward a steady state with quiescent accretion flow, as well as the Mach 3 and 10 models with an accretor size of $1R_A$. Only the models at Mach 3 and 10 and with accretor sizes 0.1 and 0.02 show unstable flow patterns.
2. A periodicity with period 0.2 time units can be discerned for the Mach 10 models.
3. The shock front is always attached to the accretor and forms a tail shock, contrary to most simulations of flow with $\gamma = 5/3$ and $\gamma = 4/3$. The fluctuations generated in the unstable models are smaller in the isothermal simulations than in the models with larger $\gamma$.
4. The smaller value of $\gamma$ (softer equation of state) has as consequence, that higher densities have to be reached before a sufficiently high pressure is built up to support the shock front (compared to models with larger $\gamma$). Thus the densities in the shock cone are higher and still the pressure is not enough to separate the shock from the accretor (tail shock). Both these consequences then produce a higher rate on linear momentum accretion and a smaller fluctuation of the cone and accretion rates.
5. The model with Mach 0.6 yields accretion rates identical (to numerical accuracy) to those in stationary accretor models.
6. The mass accretion rates are slightly larger for the $\gamma = 1.01$ models compared to $\gamma = 4/3$. The phenomenological fit we proposed in a previous paper and which worked well for the

7. The rms of the specific angular momentum accreted in the isothermal simulations is about a factor of 2–3 smaller than in the models for $\gamma = 5/3$ and $\gamma = 4/3$.
8. A comparison of the hydrodynamical drag (= accreted linear momentum) yields: (1) only the accretors with radius 1 $R_A$ are subjected to significant drag. Smaller accretors in subsonic models experience hardly any drag at all, while small accretors in supersonic models are accelerated (i.e. the relative velocity between accretor and medium is increased) by the momentum they accrete. However, the gravitational force always acts as drag and is stronger than the momentum accretion drag, so the net force is decelerating. (2) The acceleration is larger with decreasing $\gamma$.
9. A comparison of the gravitational drag (front-back density asymmetry) yields: (1) it is always positive, i.e. decelerating. (2) The subsonic models experience a drag roughly one order of magnitude smaller than the supersonic models. (3) The drag decreases with increasing accretor size, with increasing Mach number and with increasing $\gamma$.

Movies in mpeg format of the dynamical evolution of all models are available in the WWW at http://www.mpa-garching.mpg.de/~mor/bhla.html

*Acknowledgements.* I would like to thank Dr. F. Meyer, Dr. U. Anzer, Dr. T. Foglizzo and K. Schenker for carefullay reading the manuscript and improving its language and contents. The calculations were done at the Rechenzentrum Garching.

Note: obtained values by as much as 60% for $\gamma = 1.01$.